\documentclass[showpacs,preprintnumbers,two column]{revtex4}
\usepackage{amsmath}
\usepackage{graphicx}
\usepackage{dcolumn}
\usepackage{bm}

\begin{document}

\title{Quantum phase transitions of the spin-boson model within
multi-coherent-states}
\author{Liwei Duan$^{1}$, Shu He$^{1}$, and Qing-Hu Chen$^{1,2,*}$}

\address{
$^{1}$ Department of Physics, Zhejiang University, Hangzhou 310027,
 China \\
$^{2}$  Collaborative Innovation Center of Advanced Microstructures, Nanjing University, Nanjing 210093, China
 }\date{\today }

\begin{abstract}
A variational approach based on the multi-coherent-state ansatz with
asymmetric parameters is employed to study the ground state of the
spin-boson model. Without any artificial approximations except for the finite  number of
the coherent states, we find the robust Gaussian critical behavior in the
whole sub-Ohmic bath regime. The converged critical coupling strength can be estimated with the $1/N$ scaling, where $N $ is the
number of the coherent states. It is strongly demonstrated
the breakdown of the well-known quantum-to-classical mapping for $1/2<s<1$.
In addition, the entanglement entropy displays more steep jump around the
critical points for the Ohmic bath than the sub-Ohmic bath.
\end{abstract}

\maketitle

\section{Introduction}

A quantum system inevitably couples to the environmental degree of freedom,
which forms an open quantum system~\cite{open}. It is of great significance
to understand the influence of environment on the quantum system. As a
paradigmatic model to study the open quantum systems, the spin-boson model
has drawn persistent attentions~\cite{sbm}. In the spin-boson model, the
quantum system is simplified as a single spin (qubit), while the environment
is abstracted into a bosonic bath with an infinite number of modes. The
coupling between the qubit and the environment is characterized by a
spectral function $J(\omega )$ which is proportional to $\omega ^s$. The
spectral exponent $s$ varies the spin-boson model into three different
types: sub-Ohmic ($s<1$), Ohmic ($s=1$), and super-Ohmic ($s>1$).

Despite its simple form, there still exist great challenges to analyze the
spin-boson model, due to the continuous bosonic bath which leads to an
infinite number of degree of freedoms. The difficulties are aggravated by
the infrared divergence of the sub-Ohmic and Ohmic spin-boson model which
has close relations with the quantum phase transition (QPT)~~\cite
{sbm,NRG,Hur,Kopp}. Many advanced numerical approaches have been applied to
this model, such as the numerical renormalization group (NRG)~\cite
{Bulla,vojta,Tong}, quantum Monte Carlo simulation (QMC)~\cite{QMC}, sparse
polynomial space approach~\cite{SPS}, exact diagonalization in terms of
shift boson~ \cite{Zhang}, and variational matrix product state method~\cite
{VMPS,Frenzel} . It is generally accepted that there exists a second-order
QPT for sub-Ohmic baths and the Kosterlitz-Thouless QPT for the Ohmic bath.

It has been argued for a long time that the QPT of the present quantum model
is in the same universality class as the thermodynamic phase transition of
the one-dimensional Ising model with long-range interactions \cite
{open,Blote,Fisher,Vojta2012}. This quantum-to-classical correspondence was
supported by most numerical approaches, but also questioned by a Berry phase
term emerged in the action~\cite{berry,Kirchner}. Most recently, we have
developed a displaced Fock State (DFS) method \cite{DFS} to analytically study
the sub-Ohmic spin-boson model and present evidence of the Gaussian
criticality persistent in the whole sub-Ohmic bath regime.

The variational study based on the polaronic unitary transformation by
Silbey and Harris ~\cite{Silbey} has inspired a lot of studies by means of
coherent states and various extensions \cite
{zheng,Chin,Chin1,cao,mcs,ZH,duan,zhou}. Zheng~\textit{et al.} reproduced
the results of Silbey-Harris ansatz by unitary transformation without
variational procedures. The zeroth-order approximation in DFS method \cite
{DFS} can also recover the famous Silbey-Harris results. A generalized
Silbey-Harris ansatz was proposed by Chin~\textit{et al.} which correctly
describes a continuous transition with mean-field exponents for $0<s<1/2$~
\cite{Chin}. However, it failed to give reliable critical points for $%
1/2<s<1 $~\cite{ZH}.

Recently, the single coherent states ansatz~\cite{Silbey} was improved by
simply adding other coherent states on the equal footing~\cite{mcs} and by
superpositions of two degenerate single coherent states~\cite{ZH}, which are
generally termed as multi-coherent-states (MCS) ansatz in this paper.
Actually, the MCS in the single-mode version has been proposed ten years
before by Ren \textit{et al.}~\cite{tcs} independently. Bera \textit{et al.}
~\cite{mcs} have studied the novel environmental entanglement and spin
coherence in the Ohmic spin-boson model~by increasing the number of the coherent states
without much more difficulties. Very interestingly, the MCS ansatz was shown
to have fast convergence and can give results with very high accuracy. The
variational study using MCS with unconstrained parameters have not been
studied in the spin-boson model, which may hopefully shed light on the
quantum criticality of the spin-boson model.

Among all single coherent ansatz, any correlations among bosons are not
included, so in principle the non-mean--field exponent can not be given.
While the correlations among phonons should be certainly embodied in the MCS
ansatz. Recently, diagrammatic multiscale methods anchored around local
approximations, where the particle correlations are self-consistently taken
into account, indeed capture the well known non-mean-field nature of a
lattice model~\cite{Antipov}. So it is expected that MCS would give the
precise description for the quantum criticality also, which motivate the
present study for the QPT in both sub-Ohmic and Ohmic spin-boson model
within the MCS ansatz without imposing any limits on the variational
parameters.

The rest of paper is organized as follows. In Sec.~II, the spin-boson
Hamiltonian is introduced briefly. The asymmetrical MCS ansatz is proposed
in Sec.~III, and the self-consistent equations for the variational
parameters are derived. The numerical results for the order parameter and
the entanglement entropy are presented and discussed in Sec. IV, and
conclusions are given in the last section.

\section{Hamiltonian}

The spin-boson Hamiltonian can be written as
\begin{equation}
\hat{H}=-\frac{\Delta }{2}\sigma _{x}+\sum_{k}\omega _{k}b_{k}^{\dag }b_{k}+%
\frac{\sigma _{z}}{2}\sum_{k}\lambda _{k}\left( b_{k}^{\dag }+b_{k}\right) ,
\end{equation}
where $\sigma _{i}$ ($i=x,y,z$) are the Pauli matrices, $\Delta $ is the
tunneling amplitude between the spin-up state $|\uparrow \rangle $ and the
spin-down state $|\downarrow \rangle $, $b_{k}$ ($b_{k}^{\dag }$) is the
bosonic annihilation (creation) operator which can (create) a boson with
frequency $\omega _{k}$, $\lambda _{k}$ is the corresponding coupling
strength between the qubit and the bosonic bath, which is determined by the
spectral density $J(\omega )$,
\begin{equation}
J(\omega )=\sum_{k}\lambda _{k}^{2}\delta (\omega -\omega _{k})=2\alpha
\omega _{c}^{1-s}\omega ^{s}\Theta (\omega _{c}-\omega ),
\end{equation}
where $\alpha $ is a dimensionless coupling constant, $\omega _{c}$ is the
cutoff frequency which is set to be $1$ throughout this paper, $\Theta
(\omega _{c}-\omega )$ is a step function.

\section{Multi-coherent-state ansatz}

The ground-state in the generalized Silbey-Harris ansatz~\cite{Chin,Chin1}
can be written in the bases of spin-up state $|\uparrow \rangle $ and
spin-down state $|\downarrow \rangle $ as
\begin{equation}
|\Psi \rangle =\left(
\begin{array}{c}
A\exp \left[ \sum_{k=1}^{N_b}f_k\left( b_k^{\dag }-b_k\right) \right]
|0\rangle \\
B\exp \left[ \sum_{k=1}^{N_b}g_k\left( b_k^{\dag }-b_k\right) \right]
|0\rangle
\end{array}
\right) ,
\end{equation}
where $A$\ ($B$) is related to the occupation probabilities of spin-up
(spin-down) state, while $f_k$ ($g_k$) are the corresponding bosonic
displacements of the $k$th mode. It can be reduced to the original
Silbey-Harris ansatz if set $A=B$, and $f_k=-g_k$. Note that the nonlocal
correlations among phonons are not included in this ansatz, so
non-mean-field nature cannot be described in this ansatz. In the recent
analytic DFS method ~\cite{DFS}, correlations for more than one phonon can
be fully considered step by step. This is a very clean and rigourous
approach where the correlations among phonons are explicitly shown. However,
even the nearly converged results for the magnetic order parameter up to the
third-order DFS still cannot give the non-mean-field nature for $s>1/2$.
Note that in the DFS, the number of the parameters for the self-consistent
solutions, which are required in the discretization in the continuous
integral, increase exponentially with the approximation order, so it 
becomes extremely difficult to explore the further corrections.

It is very interesting to note that the correlations among more phonons can
be also included in the MCS ansatz~\cite{tcs,mcs}. More importantly, the
number of variational parameters only increases with the number of the
coherent states linearly, so the high order of approximations can be
practically performed. It is  expected that the MCS can provide insights
upon the nontrivial critical behavior of the spin-boson model.

A general form of the MCS in the bases of spin-up state $|\uparrow \rangle $
and spin-down state $|\downarrow \rangle $ can be written as
\begin{equation}
|\Psi \rangle =\left(
\begin{array}{c}
\sum_{n=1}^NA_n\exp \left[ \sum_{k=1}^{N_b}f_{n,k}\left( b_k^{\dag
}-b_k\right) \right] |0\rangle \\
\sum_{n=1}^NB_n\exp \left[ \sum_{k=1}^{N_b}g_{n,k}\left( b_k^{\dag
}-b_k\right) \right] |0\rangle
\end{array}
\right) ,  \label{GS_wave}
\end{equation}
where $A_n$\ ($B_n$) are related to the occupation probabilities of spin-up
(spin-down) state, while $f_{n,k}$ ($g_{n,k}$) are the corresponding bosonic
displacements of the $k$th mode, and $|0\rangle $\ is the vacuum state of
the bosonic bath. $N$ is the number of the coherent states, and $N_b$ is the
number of the discrete bosonic modes. A generalized Silbey-Harris ansatz
\cite{Chin} can thus be recovered if set $N=1$. The symmetric MCS ansatz ($%
A_n=B_n$ and $f_{n,k}=-g_{n,k}$) can only be applied to the delocalized
phase. In the spin-boson model, since the QPT may occur with a symmetry
breaking, the asymmetric wavefunction {\ref{GS_wave}) should be generally
employed}.

The energy expectation value can be expressed as
\begin{equation}
E=\frac{\langle \Psi |\hat{H}|\Psi \rangle }{\langle \Psi |\Psi \rangle },
\label{ehd}
\end{equation}
where
\begin{eqnarray*}
\langle \Psi |\hat{H}|\Psi \rangle  &=&\sum_{m,n}(A_mA_nF_{m,n}\alpha _{m,n}
\\
&&+B_mB_nG_{m,n}\beta _{m,n}-\Delta \Gamma _{m,n}A_mB_n), \\
\langle \Psi |\Psi \rangle  &=&\sum_{m,n}\left(
A_mA_nF_{m,n}+B_mB_nG_{m,n}\right) ,
\end{eqnarray*}
with
\begin{eqnarray*}
F_{m,n} &=&\exp \left[ -\frac 12\sum_k\left( f_{m,k}-f_{n,k}\right)
^2\right] , \\
G_{m,n} &=&\exp \left[ -\frac 12\sum_k\left( g_{m,k}-g_{n,k}\right)
^2\right] , \\
\Gamma _{m,n} &=&\exp \left[ -\frac 12\sum_k\left( f_{m,k}-g_{n,k}\right)
^2\right] , \\
\alpha _{m,n} &=&\sum_k\left[ \omega _kf_{m,k}f_{n,k}+\frac{\lambda _k}%
2\left( f_{m,k}+f_{n,k}\right) \right] , \\
\beta _{m,n} &=&\sum_k\left[ \omega _kg_{m,k}g_{n,k}-\frac{\lambda _k}%
2\left( g_{m,k}+g_{n,k}\right) \right] .
\end{eqnarray*}
The parameters $\left\{ A_n\right\} $, $\left\{ B_n\right\} $, $\left\{
f_{n,k}\right\} $, and $\left\{ g_{n,k}\right\} $ are determined by
minimizing the energy expectation value $E$ with respect to the variational
parameters. The total number of  variational parameters is $2N\left(
N_b+1\right) $. For $A_n$ and $B_n$, we have
\begin{eqnarray}
\sum_n\left( 2A_nF_{i,n}\left( \alpha _{i,n}-E\right) -\Gamma
_{i,n}B_n\Delta \right)  &=&0,  \label{aa} \\
\sum_n\left( 2B_nG_{i,n}\left( \beta _{i,n}-E\right) -\Gamma _{n,i}A_n\Delta
\right)  &=&0,  \label{bb}
\end{eqnarray}
and for $f_{n,k}$ and $g_{n,k}$, we obtain
\begin{eqnarray}
&&\sum_n\{-\Delta \Gamma _{i,n}B_ng_{n,k}  \nonumber  \label{ff} \\
&&+A_nF_{i,n}\left[ 2\left( \alpha _{i,n}+\omega _k-E\right) f_{n,k}+\lambda
_k\right] \}=0, \\
&&\sum_n\{-\Delta \Gamma _{n,i}A_nf_{n,k}  \nonumber \\
&&+B_nG_{i,n}\left[ 2\left( \beta _{i,n}+\omega _k-E\right) g_{n,k}-\lambda
_k\right] \}=0.  \label{gg}
\end{eqnarray}
In practise, these parameters can be obtained by solving the coupled
equations self-consistently, which in turn give the GS energy and
wavefunction. Generally, the largest number of the coherent states in the
practical calculations  can be reached up to  $N_{\max }=10$  in this paper.

In the DFS study ~\cite{DFS}, we have rigourously proved that all summation
over $k$ is related to $\sum_k \lambda_k^2,$ and can be transformed into the
continuous integral $\int_0^{\omega _c}d\omega J(\omega )I(\omega ).\;$To be
free from any approximation except for the finite number of the MCS, the
exact summation over $k$ should be performed. We here use the
Gaussian-logarithmical discretization developed in Ref.~\cite{DFS} to
calculate the continuous spectral numerically exactly by checking the
convergence. The widely used logarithmical discretization usually
overestimate the critical points~\cite{DFS}.

\begin{figure}[tbp]
\includegraphics[width=8cm]{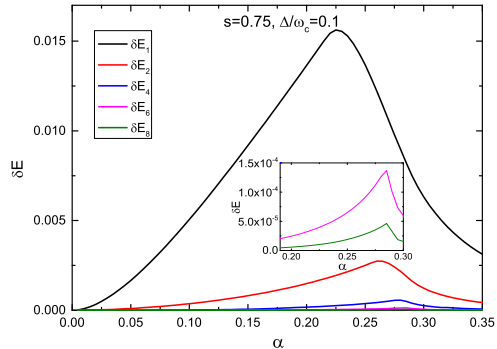}
\caption{ (Color online) The ground state energy relative difference $\delta
E_N=(E_{N}-E_{10})/E_{10}$ as a function of $\alpha$ for various numbers of
coherent states. The inset is a enlarged view for $N=6$ and $8$ curves. $%
s=0.75 $ and $\Delta/\omega _{c}=0.1$. }
\label{energy}
\end{figure}

In order to check the validity of the MCS ansatz with unconstrained the
parameters, we compare the ground state energy calculated by Eq.~(\ref{ehd})
for the different number of coherent states ($N$). We present the relative
difference $\delta E_N=\left( E_N-E_{N_{\max }}\right) /E_{N_{\max }}$ as a
function of the coupling strength $\alpha $ in Fig.~\ref{energy} for $s=0.75$
and $\Delta /\omega _c=0.1$. It is shown that $\delta E_N$ becomes smaller
rapidly with increasing $N$. $\delta E_{10}$ is always less than $10^{-5}$ in the whole
coupling regime, demonstrating that an excellent convergence is achieved for
$N_{\max }=10$. In other words, $N_{\max }=10$ is sufficient large to
describe the ground-state in this model. In the next section, we will study
the criticality of the spin-boson model with the MCS by using the
numerically exact Gaussian-logarithmical integrations.

\section{Results and discussions}

\begin{figure}[tbp]
\includegraphics[width=8cm]{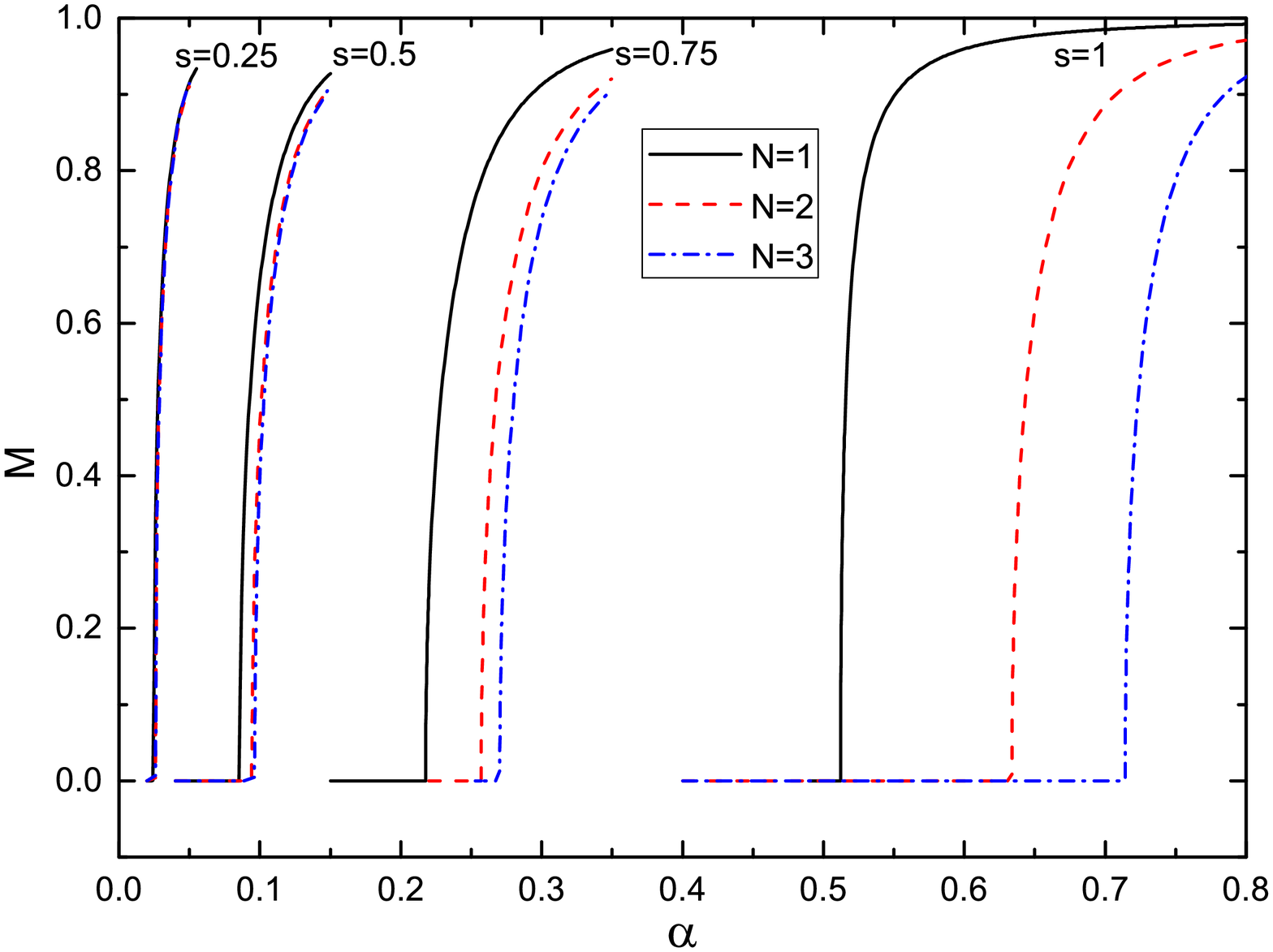}
\caption{(Color online) The magnetization $M$ as a function of $\alpha$ for $%
s=0.25$, $0.5$, $0.75$, and $1$ with  $N=1,
2 $, and $3$. $\Delta /\omega _c=0.1$.}
\label{pz}
\end{figure}

It is well-known that the magnetization $M$ acts as an order parameter in
the sub-Ohmic spin-boson model, which can be written as
\begin{equation}
M=\frac{\langle \Psi |\sigma _z|\Psi \rangle }{\langle \Psi |\Psi \rangle }%
=\frac 1D\sum_{m,n}\left( A_mA_nF_{m,n}-B_mB_nG_{m,n}\right) .
\end{equation}
Figure.~\ref{pz} shows the value of   $M$ as a function of $\alpha $ for
different numbers of coherent states. It is obvious that there exists a
critical coupling strength $\alpha _c$ which separates the delocalized phase
with zero magnetization from the localized one with nonzero magnetization.
In the weak coupling regime, the magnetization $M$ is always zero,
indicating that the spin  stay in the spin-up and the spin-down states with
the same probability. With increasing coupling strength,  $M$ will tend to  $%
1$ (due to the degeneracy, another branch will tend to  $-1$ symmetrically).
The critical coupling strength $\alpha _c$ increases with the value of  $s$,
as shown in Fig.~\ref{pz}.  For same spectral exponent $s$, the obtained $%
\alpha _c$ increases with the number of coherent state. The Ohmic case was
studied by the symmetric MCS ansatz for $0<\alpha <1$ in Ref.~\cite{mcs}. It
should be noted that the symmetric MCS is invalid in the localized phase.

\begin{figure}[tbp]
\includegraphics[width=8cm]{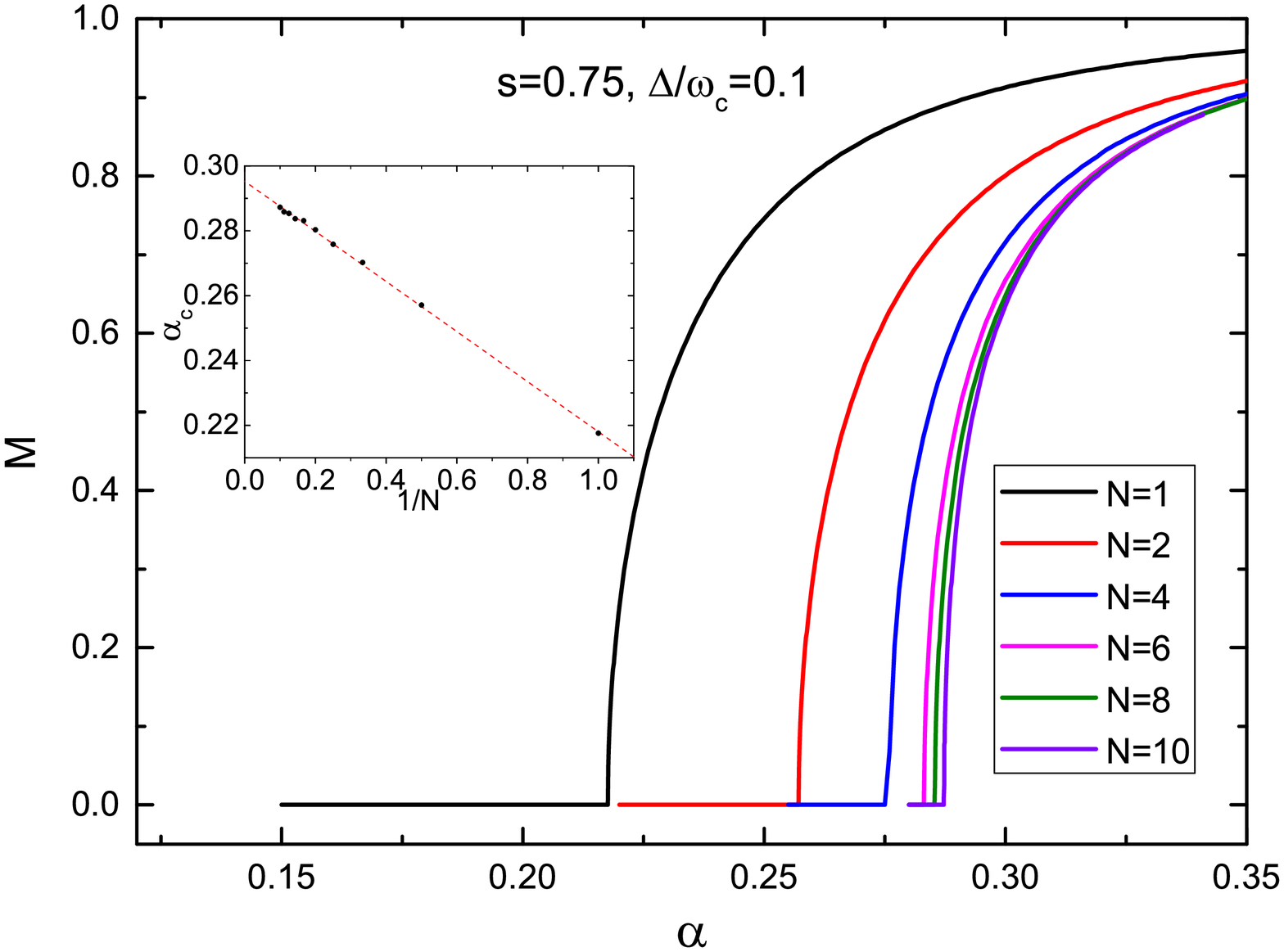}
\caption{(Color online) The magnetization $M$ as a function of $\alpha $
with seven numbers of coherent states for $s=0.75$ and $\Delta /\omega _c=0.1
$. Inset shows the critical coupling strength $\alpha _c$ as a function of
the inverse coherent state number $1/N$. A very nice linear fitting yields
the extrapolated limiting value $\alpha _c=0.2952$ as $N\rightarrow \infty $%
. }
\label{s075}
\end{figure}

We then pay particular attention to $s=0.75$, which is typical value greater
than $0.5$. As shown in Fig.~\ref{s075}, a fast convergence for the
magnetization can be achieved with increasing number of the coherent states.
In principle, the true critical coupling strength $\alpha _c$ should only be
obtained by the infinite number of the coherent states in the MCS, which
definitely cannot be realized in the practical calculations. It also
becomes extremely difficult to solve the self-consistent equations~[\ref{aa}%
]-[\ref{gg}] for an unbounded large number of coherent states, especially
near critical points. Fortunately, a perfect linear scaling $\alpha _c$ as
a function of the inverse coherent state number $1/N$ is shown in the
inset of Fig.~\ref{s075}. In this way, $\alpha _c$ is extrapolated to $%
0.2952 $ when $1/N\rightarrow 0$, i.e. $N\rightarrow \infty $. Very
interestingly, this value for $\alpha _c$ is consistent excellently with $%
0.2951$ obtained by QMC~\cite{QMC}, the only numerical method where the
discretization of the bath is not needed in literature, to the best of our knowledge.

The magnetization shows a power law behavior near the critical point, namely
$M\propto (\alpha -\alpha _c)^\beta $. The classical counterpart of the
sub-Ohmic spin-boson model is the one-dimensional Ising model with
long-range interaction according to the quantum-to-classical mapping~\cite
{open,Blote,Fisher}. It is predicted that a continuous phase transition with
mean-field behavior undergoes for $0<s<1/2$ and non-mean-field behavior for $%
1/2<s<1$. Most previous numerical approaches demonstrated the validity of
the quantum-to-classical mapping. Surprisingly, the DFS method \cite{DFS} gives the robust mean-field exponent $%
\beta =1/2$ in the whole sub-Ohmic regime $0<s<1$, which is in sharp
contrast with the previous conclusions. Therefore, the critical behavior of
the sub-Ohmic spin-boson model needs further extensive direct studies where
the Berry phase and topological effects are not missed~\cite{berry,Kirchner}%
. The studies based on the Feynman path-integral representation of the
partition function may not satisfy this requirement.

\begin{figure}[tbp]
\includegraphics[width=8cm]{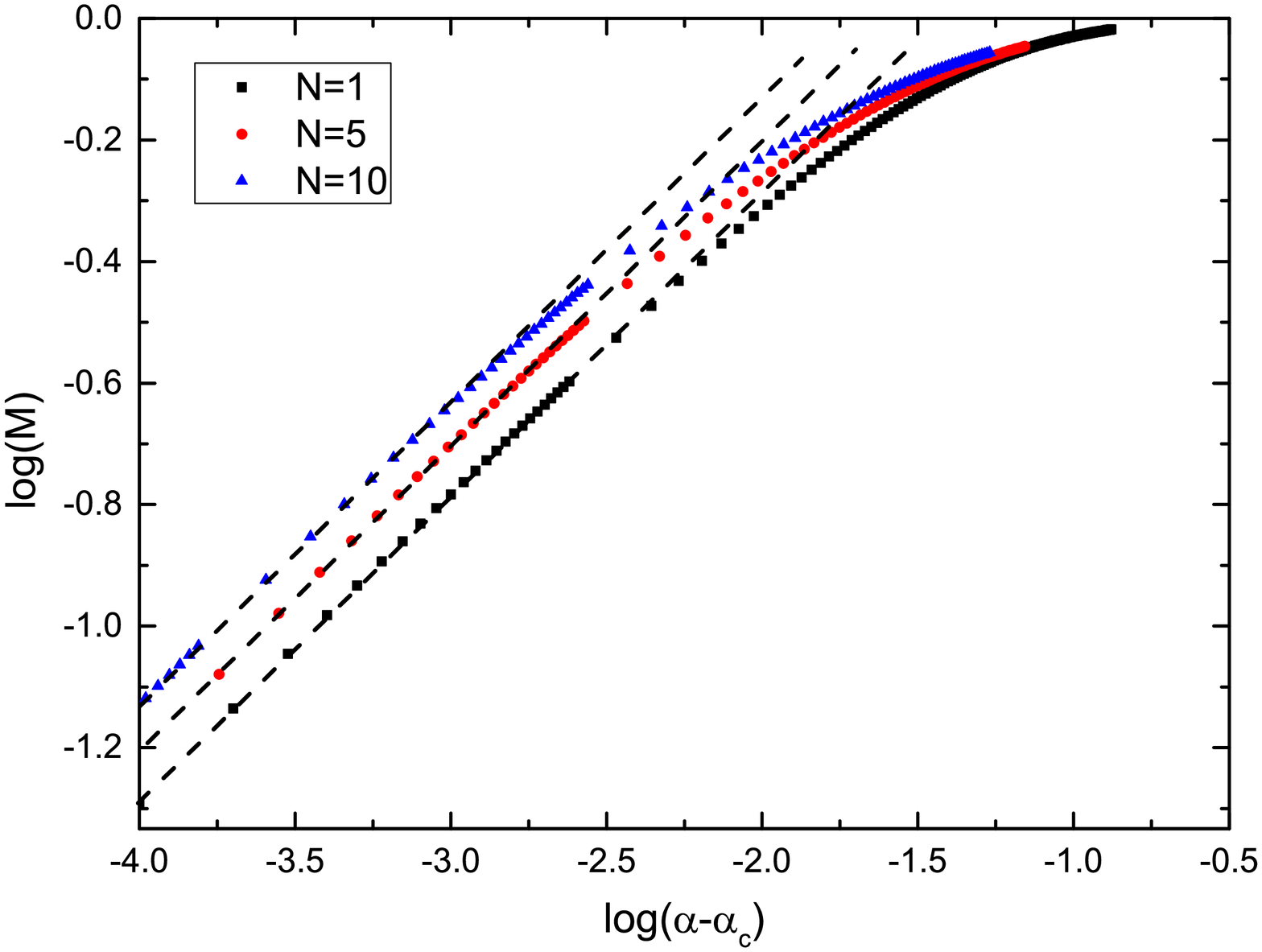}
\caption{(Color online) Log-log plot of magnetization $M$ as a function of $%
(\alpha -\alpha _c)$ for $s=0.75$ and $\Delta /\omega _c=0.1$ with the
number of coherent state $N=1$ (square), $5$ (cycle) and $10$ (triangle).
The power law curves with $\beta=0.5$ is denoted by the dashed line. }
\label{criticalE}
\end{figure}

\begin{figure}[tbp]
\includegraphics[width=15cm]{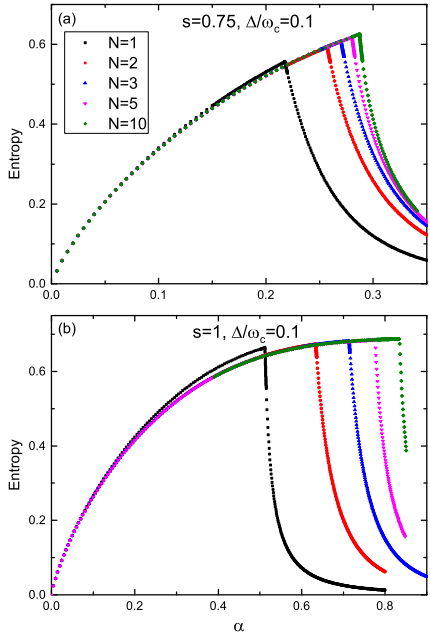}
\caption{(Color online) The entanglement entropy as a function of the
coupling strength for $s=0.75$ (a) and $s=1$ (b).}
\label{entanglement}
\end{figure}

Figure.~\ref{criticalE} displays the log-log plot of magnetization as a
function of $(\alpha -\alpha _c)$ for $s=0.75$ for $N=1, 5$, and $10$.
 It is surprising to observe that all curves show a very
nice power-law behavior over more than $2$ decades with an exponent $\beta
=0.5$. It is strongly suggested that even   $N=10$ coherent states can hardly
 modify the exponent $\beta $. In other words, the number of
coherent state $N$ has negligible effect on the critical exponent $\beta
\sim 1/2$, unlike the critical coupling strength $\alpha _c$ which is quite
sensitive on $N$. So here we provide another piece of  evidence for the
breakdown of the quantum-to-classical mapping for $1/2<s<1$, besides the DFS
study~\cite{DFS}.

Now we move to the Ohmic bath case ($s=1$). To demonstrate the difference
between the $s=1$ and $s<1$ baths, we  study the entanglement entropy
between the qubit and the bath. In the spin-boson model, entanglement
entropy can be obtained as~\cite{Kopp}
\[
S=-p_{+}\log p_{+}-p_{-}\log p_{-},
\]
where $p_{\pm }=\left( 1\pm \sqrt{\langle \sigma _x\rangle ^2+\langle \sigma
_z\rangle ^2}\right) /2.\;$In the exact NRG study \cite{Hur}, $-\langle
\sigma _x\rangle =\Delta /\omega _c$ for $\alpha \rightarrow 1$. So in the
delocalized phase, the entailment entropy should converge to $S\simeq \ln 2-%
\frac{\langle \sigma _x\rangle ^2}{\ln 10}$ at the strong coupling, which is
$0.\,6888$ for $\Delta /\omega _c=0.1$.

The entanglement entropy as a function of the coupling strength is given in
Fig. \ref{entanglement} for $s=0.75$ and $s=1\;$at $\Delta /\omega _c=0.1$.
We observe that the entanglement entropy exhibits a cusp for $s<1$, and shows
a very steep drop for $s=1$ at the critical points. With increase of the
number of the coherent states, the entanglement entropy jump more steeply at
the critical point. Interestingly, the entanglement entropy curves for $s=1$
resemble very much curves for unbiased case in Fig. 4 of Ref. \cite{Hur} by
the Bethe ansatz study at extremely low temperature. It is expected that the
entanglement entropy will vertically jump to zero in the limit of $%
N\rightarrow \infty $. This may be a signature of the Kosterlitz-Thouless
type QPT from a second-order QPT. The Kosterlitz-Thouless phase transition,
of infinite order, for the Ohmic bath ($s=1$)~\cite{open,sbm}, can be only
reached in the infinite number of coherent states. Such a sudden jump is
not, but perhaps related to the sudden jump of the superfluid density in
thin films of $He^4$ described by the two-dimensional XY model~\cite{Nelson}%
, which is worthy of a further study.

\section{Conclusion}

In this work, by means of the asymmetric MCS ansatz, we extensively analyze
the ground state of the spin-boson model, especially the quantum
criticality. This direct study to the quantum model does not miss the Berry
phase or topological defects. Without any artificial approximations except
for the finite number of the coherent states, we find that the magnetic
exponent $\beta \sim 1/2$ at the critical points is robust in the whole
sub-Ohmic bath regime, consistent with the recent DFS study and  in
contrast with most numerical studies in literature. It is strongly suggested
that the well-known quantum-to-classical mapping is broken down for $1/2<s<1$
. The asymptotical behavior in the infinite number of the coherent states
are also analyzed. The converged critical strengths for $s=0.75\ $ agrees
well with the QMC formulated on Feynman path-integral representation of the
partition function. For $s=1$, more steep jump of the entanglement entropy
at the critical point for the larger number $N$ of the coherent states is
observed. It is expected that the vertical jump of the entanglement entropy
would occur at the Kosterlitz-Thouless phase transition points in the large $%
N$ limit.

\acknowledgments  This work is supported by National Natural Science
Foundation of China under Grant No. 11474256, and National Basic Research
Program of China under Grant No.~2011CBA00103.\

$^{\ast }$ Corresponding author. Email:qhchen@zju.edu.cn

\end{document}